\documentclass[10pt]{article}

\usepackage[english]{babel}
\usepackage{graphics,graphicx}
\usepackage{amssymb,amsfonts}
\usepackage{amsmath}
\usepackage[pdfstartview=FitH,colorlinks,linktocpage,unicode]{hyperref}
\usepackage{cite}
\usepackage{indentfirst}
\usepackage{color}

\newcommand{\beq}{\begin{eqnarray}}
\newcommand{\eeq}{\end{eqnarray}}

\newcommand{\const}{\text{const}}

\begin{document}
\title{Emergence of symmetries}
\author{A.~A.~Kirillov\thanks{kirillov-aa@yandex.ru} \and A.~A.~Korotkevich\thanks{korotkevich\_aa@mail.ru} \and S.~G.~Rubin\thanks{sergeirubin@list.ru} \\National Research Nuclear University ``MEPhI''}
%
%



\date{\today}

\maketitle

\begin{abstract}
The mechanism of symmetry formation is discussed in the framework of multidimensional gravity. It is shown that this process is strictly connected to the entropy decrease of compact space. The existence of low energy symmetries is not postulated from the beginning. They could be absent during the inflationary stage under certain conditions discussed in the paper.
\end{abstract}

\section{Introduction}
The idea of multidimensional space-time allows to clarify some fundamental questions, such as the problems of modern cosmology and the Standard Model which are discussed in terms of extra-dimensional gravity \cite{1998PhLB..429..263A,1999NuPhB.537...47D, 1999PhRvL..83.3370R, 2002PhRvD..65b4032A,2004PhRvD..70a5012C,2005PhRvD..71c5015C,2007CQGra..24.1261B}. As was shown in \cite{2009JETP..109..961R}, the numerical values of the fundamental parameters depend on geometry of extra dimensions. The existence of gauge symmetries may be related to isometries of extra space \cite{Blagojevic,2008IJMPD..17..785C,2009JETP..109..961R}. Hereinafter we consider compact d-dimensional extra spaces in the spirit of the Universal Extra Dimensions (UED). It is supposed that symmetries of extra space are related to the low energy symmetries in the observable universe. The size of extra dimensions varies in the wide range from the Planck scale up to $10^{-18}$cm, the upper limit known from the particle physics. This approach is similar to the UED one though no fields except metric tensor are considered. An applicability of the results discussed below to another approaches like ADD  \cite{1998PhLB..429..263A} and Randall-Sundrum models \cite{1999PhRvL..83.3370R} is the subject of future investigations.

(Maximally) symmetric metrics of extra space as a starting point are among the most popular in the literature, see e.g. \cite{2008IJMPD..17..785C,2011PhR...497...85M,2011PhRvD..84d4015B}. This assumption makes it possible to obtain clear and valuable results.
At the same time we must take into account the quantum origin of space itself due to fluctuations in the space-time foam. There is no reason to assume that the geometry or/and topology of extra space is simple \cite{1991PhLB..259...38L,Lindebook}. Moreover it seems obvious that a measure $\mathfrak{M}$ of all symmetrical spaces equals zero so that the probability of their nucleation $\mathcal P=0$. Hence some period of extra space symmetrization ought to exist \cite{2001PhRvD..63j3511S,2009JETP..109..961R,2002PhRvD..66b4036C,2003Ap&SS.283..679G,2006PhRvD..73l4019B}.

In the present paper we investigate the entropic mechanism of space symmetrization after its nucleation. It is shown that the stabilization of the extra space and its symmetrization are proceeding simultaneously. This process is accompanied by a decrease in entropy for the extra space and an increase in entropy for main one.

\section{Time dependence of compact space geometry}

As was mentioned above some mechanism of the extra space
symmetrization should exist. In this Section we consider some toy
models to clarify the situation.

As a common basis, consider a Riemannian manifold
\begin{equation}
 T\times M\times M'
  \label{MD}
\end{equation}
with the metric
\begin{equation}
  ds^2 = G_{AB}dX^{A}dX^{B}=dt^2 - g_{mn}(t,x)dx^{m}dx^{n}-\gamma_{ab}(t,x,y)dy^{a}dy^{b}.
  \label{interval}
\end{equation}
Here $M$, $M'$ are the manifolds with spacelike metrics
$g_{mn}(t,x)$ and $\gamma_{ab}(t,x,y)$ respectively, $T$ denotes
the timelike direction.  The set of coordinates of the subspaces
$M$ is denoted by $x$; $y$ is the same for $M'$. We will refer to
$M$ and $M'$ as a main space and a compact extra space respectively. The
curvature of the manifold is assumed to be arbitrary.

Firstly, consider the ($d+1$)-dimensional compact manifold
$M'\times T$ with metric
$$
  ds^2 = dt^2 -\gamma_{ab}(t,y)dy^{a}dy^{b} ; \quad\gamma_{ab}(y,t)=\eta_{ab} + h_{ab}(t,y) .
$$
For the Einstein-Hilbert action
%
\begin{equation}
  \label{R}
  S=\int d^d y dt \sqrt{|\gamma|}R
\end{equation}
and in the limit $h_{ab}\ll 1,$ classical equations have the form
\cite{Blagojevic} \beq
\Box_{d+1} h_{ab}=0,
\eeq
where
\begin{equation}\label{Dalambe}
  \Box_{d+1}\equiv \frac{1}{\sqrt{\gamma}}\partial_0 (\sqrt{\gamma}\partial_0 ) +
  \frac{1}{\sqrt{\gamma}}\partial_a (\sqrt{\gamma}\gamma^{ab}\partial_b ).
\end{equation}
This wave equation has no static symmetrical solutions if initial conditions are arbitrary. This would mean the absence of symmetries even in the modern epoch, which is unacceptable.

The situation changes considerably if we take into account the
dynamics of the main manifold $M$. Let it possess the
Friedmann-Robertson-Walker (FRW) metric and the scale factor
$a(t)$ (we assume $\dot a(t)>0$). If our Universe plays the role of the manifold $M$ the latter statement is just the observable fact. The equation of motion for the metrics of the extra space $M'$ acquires the form
\begin{equation}
  \label{FricLin}
  \Box_{d+1}h_{ab} +3H\dot{h}_{ab}=0,
\end{equation}
where the Hubble parameter $H=\dot{a}/a >0$. We also took into account the form of metrics \eqref{interval} and equality
\begin{equation}\label{FRW}
  \frac{1}{\sqrt{g}}\partial_0 \sqrt{g} = 3\frac{\dot{a}}{a}=3H >0
\end{equation}
valid for 4-dimensional FRW space. The
term $3H\dot{h}_{ab}$ in \eqref{FricLin} indicates the presence of
friction and gives the asymptotic $\gamma_{ab}\rightarrow \const$
for $t\rightarrow +\infty$.

So the dynamics of the main space $M$ could cause the stabilization of the extra space $M'$. Note that friction usually means entropy increasing in any system.

As a more complex  and valuable example consider a gravity with higher order derivatives and the action in the form
\begin{equation}
  \label{fR}
  S=\int d^{D+1} z \sqrt{G}f(R),
\end{equation}
where $z=(t,x,y)$ and $G=|\det g\cdot \det\gamma|$. The metric of extra space \eqref{interval} is chosen in the form
$\gamma_{ab}=\gamma_{ab}(t,y)$. We also use
inequality
\begin{equation}
  \label{cond3}
  R_M \ll R_{M'}
\end{equation}
for the Ricci scalar of the main space $R_M$ and the Ricci scalar of the extra space $R_{M'}$. It is known that in the framework of D-dimensional gravity linear in the Ricci scalar the stabilization of an extra space is impossible without involving additional fields \cite{1991PhRvD..43..995H,1997PhRvD..55.6092M,2002PhRvD..66b4036C}. On the other side a D-dimensional gravity with higher derivatives gives such an opportunity. The process of stabilization of the extra space is discussed in the papers  \cite{2005CQGra..22.3135G,2006PhRvD..73l4019B} where it was shown that a stationary size of extra space depends on its dimensionality and initial parameters of the pure gravitational lagrangian.

Recall that action \eqref{fR} is equivalent to linear action with an additional scalar field \cite{1988PhLB..214..515B,1989PhRvD..39.3159M, 2005CQGra..22.3135G}
\begin{equation}
\label{Rphi}
S=\int d^{D+1} z \sqrt{\tilde{G}}\left[ \tilde{R}(\tilde{G})+\tilde{G}^{ab}\partial_a\phi\partial_b\phi -2U(\phi)\right],
\end{equation}
where
\begin{eqnarray}\label{detphi}
&&\phi=\frac{1}{A}\ln f'(R) ;\quad A=\sqrt{\frac{D-1}{D}} \label{phiR} \\
&&U(\phi)=\frac{1}{2}e^{-B\phi}\left[ R(\phi)e^{A\phi}-f(R(\phi))\right], \quad B=\frac{D+1}{\sqrt{(D-1)D}} \label{potential}
\end{eqnarray}
The details can be found in the papers cited above. The classical equation of motion of action \eqref{Rphi} has the form
\begin{equation}\label{eqphi}
\ddot{\phi}+3H\dot{\phi}+ \Box_{d+1}\phi +U'(\phi)=0,
\end{equation}
where metric \eqref{interval}, equality \eqref{FRW} and definition \eqref{Dalambe} are taken into account.
As in the previous cases, the term containing the Hubble parameter $H$ is responsible for friction in the system.

Let the potential \eqref{potential} has a minimum at $\phi_m$.
Due to the presence of the friction, the additional scalar field $\phi$ tends to a constant. According to (\ref{phiR}) the Ricci scalar of the extra space is connected to the scalar field and also tends to a stationary value,
\begin{equation}\label{Rd}
 R\rightarrow R(\phi_m)
\end{equation}
The observed main space is described by the FRW metric and its Ricci scalar tends to zero, $R_M \sim 1/a(t\rightarrow \infty)^2 \rightarrow 0$   so that we may neglect its contribution at large times. Thus the extra space $M'$ acquires a maximally symmetrical form.

As in the previous case, see \eqref{FricLin}, it is the dynamics of the main space that is responsible for the friction in the extra space and its stabilization. This indicates the presence of entropy flow from the extra space $M'$ to the main one $M$ \cite{1983PhRvL..51..931A,1984PhRvD..30.1205K}.

\section{Entropy and symmetry formation}

In the previous section we saw that stabilization of extra space, an extension of its symmetry group  and an entropy increase in a whole space proceed simultaneously. Below we show that the entropy of a compact extra space is decreasing with time. To this end we prove the following

\medskip

\emph{Statement}

\medskip

\emph{Let $M$ be a smooth manifold, $G_1$ and $G_2$ are two given metrics on it. If the number of Killing vectors of metric $G_1$ is less then the number of Killing vectors of metric $G_2$ then the entropy of $G_1$ is greater than the entropy of $G_2$.}

\medskip

We will use the well known definition of the Boltzmann entropy $S$. It links entropy to a number of microstates $\Omega$, $S =k_B \ln \Omega $. Other definitions are discussed in \cite{1999PhRvE..59.6370B,Katok} for example.

Let us consider a compact smooth manifold $M$. We suppose that every metric $G$ on $M$ defines a microstate. More definitely, two metrics $G_1$ and $G_2$ on $M$ define the same microstate if and only if they are equal in each of the points $P \in M$.

The definition of a macrostate is as follows. Let $v$ be an
arbitrary smooth vector field defined globally on the smooth
manifold $M$. Any shift along the integral path of vector field
$v$ corresponds to a diffeomorphism $M$ on itself. We define a
macrostate as a set of metrics $G$ that are connected by shifts. As
an example, a 2-dim torus with a bulge, being shifted, still
represents the same macrostate. Another macrostate is determined
by the addition of another bulge. So this definition seems
reasonable.

The statistical weight of a given macrostate is the number of
microstates. The latter is a continuum set for any classical
system. The concept of microstates is correctly defined at a quantum level where the set of energy levels is known. However, the quantization of geometry is a yet unsolved problem. That is why any discussion of a metric on scale less than the Planck scale $L_P$ is pointless. Thus shifts less than Planck scale should not be taken into account when counting statistical
weight (see discussion in \cite{2007CQGra..24..243C}). Therefore a number of shifts along various integral paths is assumed to be finite.

Let us compare statistical weights of two metrics $G_1$ and $G_2$ with the same number of shifts at manifold $M$. Let $G_1$
have no Killing vectors and $G_2$ possesses a global Killing field. Shifts along Killing vector of $G_2$ lead to the same
microstate by definition. So the statistical weight of $G_1$ is greater than the statistical weight of $G_2$. A similar argument is correct in the general case as well when the number of Killing vectors of metrics $G_1$ is less then the number of Killing vectors of metrics $G_2$. This statement is also valid for the entropy which is the nondecreasing function of the statistical weight. Therefore
\begin{equation}\label{SAB}
S_1 > S_2.
\end{equation}
The statement is proved.

So the entropy decreases in the presence of Killing vectors.
Similar result was discussed in the framework of the topological entropy \cite{Katok}. Namely if $M$ is a smooth manifold with some metric $g$ and $f$ is a shift along its Killing vector then the topological entropy $h_g(f)=0$.

Combining the results discussed above we can conclude that the entropy of the whole manifold is increasing with time, while the extra space metric tends to maximally symmetric one. The latter means an entropy decreasing of the extra space. So an isometry group of extra space becomes larger due to an influx of entropy into a main space.

The entropic flux can be realized by various mechanisms. In this section we show that the space expansion is responsible for the friction what leads to the attenuation of motion in a compact extra space. The friction term is proportional to the Hubble parameter so that most intensive flux takes place during the inflationary stage.

Another way to support the entropic flux is the decay of excitations of the compact extra space, the Kaluza-Klein modes. The dynamical evolution of compact hyperbolic extra space leads to an intensive injection of entropy into the observable Universe \cite{2001PhRvD..63j3511S,2001PhRvL..87w1303S}.

The authors of the paper \cite{2006PhRvD..73h6002B} studied an interaction of radion, matter and metric fluctuations  in the Universe that exhibits a transient stabilizing epoch of extra space.

\section{Consequences}
According to condition \eqref{Rd} a Ricci scalar of internal metric of extra space is a constant with good accuracy whenever condition \eqref{cond3} holds. Meantime the latter can be violated at the inflationary stage due to metric fluctuations of the main space.

Let us estimate the characteristic size $L_{\rm extra}$ of extra space, insensitive to this fluctuations. During the inflationary stage we have \cite{Lindebook}
\begin{equation*}
    R_3 = 12 H^2 =\frac{32\pi}{M_{\rm Pl}^2} V(\phi) = 16\pi \left(\frac{m}{M_{\rm Pl}}\right)^2\phi^2.
\end{equation*}
The last equality holds for a quadratic inflaton potential $V(\phi)=\frac12 m^2\phi^2$. Metric fluctuations of the main space have the form
\begin{equation*}
    \delta R_3 = 32\pi \left(\frac{m}{M_{\rm Pl}}\right)^2\phi\;\delta \phi \sim \frac{m}{M_{\rm Pl}}m^2.
\end{equation*}
Here we take into account the approximate equality $\phi \sim M_{\rm Pl}$ during the inflationary stage. Field fluctuations $|\delta\phi|=H/(2\pi)$ are connected to the scale factor
\begin{equation*}
    H = \sqrt{\frac{8\pi}{3}\frac{V(\phi)}{M_{\rm Pl}^2}} = \sqrt{\frac{4\pi}{3}}\frac{m}{M_{\rm Pl}}\phi
\end{equation*}
in the usual manner.
For the size of extra space not to be disturbed, it should satisfy the inequality
\begin{equation}\label{estim}
 L_{\rm extra}\sim \frac{1}{\sqrt{R_{M'}}} < \frac{1}{\sqrt{\delta R_3}} \sim \frac{1}{m}\sqrt{\frac{M_{\rm Pl}}{m}}\sim 10^{-24} \text{\ cm}.
\end{equation}
Otherwise metric fluctuations of the main space would influence the geometry of the extra space and any symmetries would be absent during inflation.

Let us suppose that a relaxation time $t_{\rm rel}$ of dynamical processes in the extra space is less than the period of inflation $t_{\rm infl}\approx 10^{-37}$, one can obtain the estimation
\begin{equation*}
 L_{\rm extra}\sim t_{\rm rel}< t_{\rm infl}\sim 10^{-27} \text{cm}.
\end{equation*}
LHC collider could find extra space provided its size is larger than $\sim 10^{-18}$ cm. Suppose that the LHC succeeded in finding an extra space.  In the framework of our approach it would mean the absence of symmetries at the inflationary stage. The same can be said about the gauge symmetries provided the gauge fields are connected to off diagonal components of metric tensor in the spirit of the Kaluza-Klein model. These symmetries arise during the stage of reheating or later. Contradiction with observations would imply serious difficulties in implementation of this approach. In this case, one should pay attention to other opportunities. For example, the same Einstein's General Theory of Relativity can be developed as the gauge theory \cite{2006PhRvD..73f5027K}.

\section{Discussion}

It is known that the idea of extra space leads to a set of observational effects. Most promising are those extra spaces which possessing some symmetries, but they are hardly produced from space-time foam. We elaborated the mechanism for the symmetries formation related to the entropy flow from the extra space to the main one. Due to the entropy decreasing in the compact subspace, its metric undergoes the process of symmetrization during some time after its quantum nucleation.
Relaxation time of symmetry restoration depends on many aspects and could overcome the period of inflation.

One can reasonably suppose that the entropy of the extra space decreases until a widest symmetry is restored. On the other side we need specific isometries to explain observable symmetries of low energy physics, $SU(2)\times U(1)$ for example. Could they be represented by a widest symmetry mentioned above? The answer is not evident though the result of the paper \cite{1987TAMS..301.413} is rather promising: every compact Lie group can be realized as the full isometry group of a compact, connected, smooth Riemannian manifold. Nevertheless this problem needs further investigation.

The authors are grateful to V. Berezin, S. Bolokhov, V. Dokuchaev and Yu. Eroshenko for fruitful discussions and valuable comments. This work was supported in part by grant 14.A18.21.0789 of the Ministry of Education and Science of Russian Federation.

\bibliographystyle{unsrt}
\bibliography{Article}

\end{document}